\documentclass[aps,prd,showpacs,superscriptaddress,nofootinbib,floatfix]{revtex4}
\usepackage{amssymb,amsmath,epsf,epsfig,graphicx}
\usepackage{color}
\newcommand{\feyn}[1]{#1\kern-0.45em/}      
\begin{document}
\title{Next-to-leading-order time-like pion form factors in $k_T$ factorization}

\author{Hao-Chung Hu}       \email{hchu@phys.sinica.edu.tw}
\affiliation{Institute of Physics, Academia Sinica, Taipei, Taiwan 115, ROC}
\author{Hsiang-nan Li}        \email{hnli@phys.sinica.edu.tw}
\affiliation{Institute of Physics, Academia Sinica, Taipei, Taiwan 115, ROC}
\affiliation{Department of Physics, National Cheng-Kung University, Tainan, Taiwan 701, ROC}
\affiliation{Department of Physics, National Tsing-Hua University, Hsinchu, Taiwan 300, ROC}

\begin{abstract}
We calculate the time-like pion-photon transition form factor and
the pion electromagnetic form factor up to next-to-leading order
(NLO) of the strong coupling constant in the leading-twist $k_T$
factorization formalism.
It is found that the NLO corrections to the magnitude (phase) are
lower than $30\%$ ($30^\circ$) for the former, and lower than $25\%$
($10^\circ$) for the latter at large invariant mass squared $Q^2>30$
GeV$^2$ of the virtual photons.
The increase of the strong phases with $Q^2$ is obtained,
consistent with the tendency indicated by experimental data. This
behavior is attributed to the inclusion of parton transverse momenta $k_T$,
implying that the $k_T$ factorization 
is an appropriate framework
for analyzing complex time-like form factors. Potential extensions of
our formalism to two-body and three-body hadronic $B$ meson decays
are pointed out.
\end{abstract}
\maketitle

\section{Introduction}

The $k_T$ factorization 
formalism \cite{CCH,CE,LRS,BS,LS,HS} has been
applied to next-to-leading-order (NLO) analysis of several
space-like form factors, such as the pion-photon transition form factor
\cite{Nandi:2007qx,Li:2009pr}, the pion electromagnetic (EM) form
factor \cite{Li:2010nn}, and the $B\to\pi$ transition form factors
\cite{Li:2012nk}. The calculations are nontrivial, because partons
off-shell by $k_T^2$ are considered in both QCD quark diagrams and
effective diagrams for meson wave functions. The gauge invariance of
hard kernels, derived from the difference of the above two sets of
diagrams, needs to be verified. The regularization of the light-cone
singularity in the effective diagrams generates double logarithms,
which should be summed to all orders. It has been found that the NLO
corrections, after the above treatments, are negligible in the pion
transition form factor, but reach 30\% in the latter two cases.
In this Letter we shall extend the NLO $k_T$ factorization formalism
to the time-like pion transition and EM form factors.

One of the widely adopted theoretical frameworks for two-body
hadronic $B$ meson decays is the perturbative QCD (PQCD) approach
\cite{KLS} based on the $k_T$ factorization. 
It has been shown that factorizable contributions to these decays can be
computed in PQCD without the ambiguity from the end-point
singularity. These computations indicated that sizable strong phases
are produced from penguin annihilation amplitudes, with which the
direct CP asymmetry in the $B^\pm\to K^\pm \pi^\mp$ decays was
successfully predicted. It is then a concern whether PQCD
predictions for strong phases are stable against radiative
corrections. The factorizable penguin annihilation amplitudes
involve time-like scalar form factors. Before completing NLO
calculations for two-body hadronic $B$ meson decays, it is possible
to acquire an answer to the above concern by studying the time-like
pion EM form factor. Besides, the PQCD formalism for three-body $B$ meson
decays \cite{CL03} has demanded the introduction of two-meson wave
functions \cite{MP}, whose parametrization also involves time-like
form factors associated with various currents. If PQCD results
for complex time-like form factors are reliable, a theoretical
framework for three-body $B$ meson decays can be constructed.

NLO corrections to time-like form factors are derived easily from
those to space-like ones by suitable analytic continuation from
$-Q^2$ to $Q^2$, with $Q^2$ denoting the momentum transfer squared.
We shall present the $k_T$ factorization formulas for the time-like
pion transition and EM form factors up to NLO at leading twist.
Following the prescription proposed in \cite{LS,KLS}, both the
renormalization and factorization scales are set to the virtuality
of internal particles. With this scale choice, it will be
demonstrated that the NLO corrections to the time-like pion
transition and EM form factors are under control at leading twist.
It implies that PQCD predictions for strong phases of factorizable
annihilation amplitudes in two-body hadronic $B$ meson decays may be
stable against radiative corrections. Moreover, we observe the
increase of the strong phases of the above form factors with $Q^2$,
consistent with the tendency indicated by experimental data. It will
be explained that this behavior is attributed to the inclusion of
the parton transverse momenta $k_T$. This consistency supports the
$k_T$ factorization 
as a potential framework for studying
complex time-like form factors and three-body $B$ meson decays.

\begin{figure}[t]
    \begin{center}
        \includegraphics[height=5.5cm]{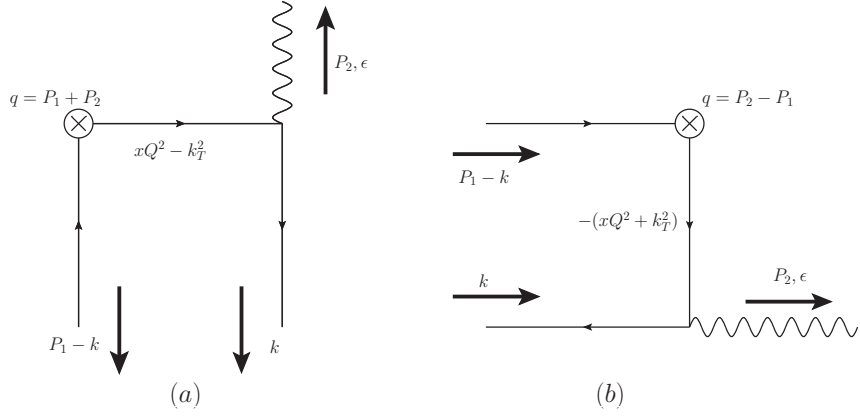}
    \end{center}
    \caption{LO quark diagrams for time-like and space-like pion-photon
    transition form factors with $\otimes$ representing the virtual photon vertex.
    The virtuality of the internal quark is labeled explicitly.}
    \label{fig:TFLO}
\end{figure}

\section{Pion-Photon Transition Form Factor}\label{PionTF}

In this section we present the leading-twist NLO factorization
formula for the time-like pion-photon transition form factor. The
leading-order (LO) QCD quark diagram describing $\gamma^*(q) \to
\pi(P_1)~\gamma(P_2)$ is displayed in Fig.~\ref{fig:TFLO}(a), where
the momentum $P_1$ of the pion and the momentum $P_2$ of the
outgoing on-shell photon are chosen as
\begin{align}
P_1 =(P_1^+,0,{\bf 0}_T) ,
\;\;\;
P_2 = (0,P_2^-,{\bf 0}_T) ,
\;\;\;
P_1^+ = P_2^- = Q/\sqrt{2} ,
\label{TFcor}
\end{align}
with $Q^2 = q^2 = (P_1 + P_2)^2 > 0$  being the invariant mass
squared of the virtual photon $\gamma^*$. Figure~\ref{fig:TFLO}(a)
leads to the LO hard kernel
\begin{align}
H^{(\text{LO})}_{\pi\gamma}(x,Q^2,k_T) =
-i\frac{N_c}{\sqrt{2N_c}}\frac{\operatorname{Tr}
[~\feyn \epsilon (\feyn P_2 + \feyn k ) \gamma_\mu \gamma^5 \feyn P_1]}
{(P_2+k)^2 + i\varepsilon}
= -i\sqrt{\frac{N_c}{2}}\frac{\operatorname{Tr}
[~\feyn \epsilon \feyn P_2 \gamma_\mu \feyn P_1\gamma^5]}
{k_T^2-xQ^2-i\varepsilon}   ,
\label{TFH0}
\end{align}
where $N_c = 3$ is the number of colors, $\epsilon$ is the
polarization vector of the outgoing photon, $k = (xP_1^+, 0, {\bf
k}_T)$ is the momentum carried by the valence quark,
$\gamma^5\feyn P_1 /\sqrt{2\text{N}_{c}}$ is the leading-twist
spin projector of the pion, and the subscript $\mu$ associated with
the virtual photon vertex is implicit on the left-hand side. In the previous works 
on the space-like transition form factor \cite{Jakob:1994hd,Li:2000hh,
Nagashima:2002ia,Nandi:2007qx}, the internal quark remains
off-shell by $(P_2-k)^2 = - (xQ^2+k_T^2)<0 $ as indicated in
Fig.~\ref{fig:TFLO}(b). For the time-like case, the internal quark
may go on mass shell, and an imaginary part is generated in the hard
kernel according to the principal-value prescription
\begin{align}
\frac{1}{k_T^2-xQ^2-i\varepsilon} = \Pr \frac{1}{k_T^2-xQ^2} +i\pi
\delta(k_T^2-xQ^2)  . \label{eq:Pri}
\end{align}

Fourier transforming Eq.~(\ref{TFH0}) into the impact-parameter $b$
space, we derive the LO pion transition form factor
\begin{align}
    F^{(\text{LO})}_{\pi\gamma}(Q^2) = i\pi\frac{\sqrt{2}f_\pi}{6} \int_0^1 dx
\int_0^\infty b db\, \phi_\pi(x)\exp[-S(x,b,Q,\mu)]\,
H^{(1)}_0\left( \sqrt{x}Qb \right)  , \label{TFI0}
\end{align}
with the pion decay constant $f_\pi$,
the renormalization and factorization scale $\mu$,
the Hankel function of the first kind $H_0^{(1)}$,
and the twist-2 pion distribution amplitude (DA) $\phi_\pi$.
Here we shall not consider the potential intrinsic $k_T$
dependence of the pion wave function \cite{Jakob},
because its inclusion would introduce additional model dependence,
which is not the focus of this work. For example, the intrinsic
$k_T$ dependence has been parameterized into the different Gaussian
and power forms in \cite{Brodsky80}.
The Sudakov factor $e^{-S}$ sums the double logarithm
$\alpha_s\ln^2 k_T$ to all orders, and takes the same expression for
both the space-like and time-like form factors \cite{Raha:2010kz},
since it is part of the universal meson wave function. For its
explicit expression, refer to \cite{Li:2001ay,LS,Li:1994zm}. Note
that Eq.~(\ref{TFI0}) can be obtained from the LO space-like pion
transition form factor in \cite{Nagashima:2002ia} by substituting $i
\pi H^{(1)}_0/2$ for the Bessel function $K_0$, as a consequence of
the analytic continuation $q^2 = -Q^2 \to (Q^2+i\varepsilon)$ in the
hard kernel.

As stated in the Introduction, the NLO hard kernel is derived by
taking the difference of the $O(\alpha_s)$ quark diagrams and the
$O(\alpha_s)$ effective diagrams for meson wave functions. The
ultraviolet divergences in loops are absorbed into the renormalized
strong coupling constant $\alpha_s(\mu)$, and the infrared
divergences are subtracted by the nonperturbative meson wave
functions. The above derivation has been demonstrated explicitly in
\cite{Nandi:2007qx} for the space-like pion transition form factor.
We repeat a similar calculation for the time-like pion transition
factor, and derive the NLO hard kernel\footnote{
    Compared to \cite{Nandi:2007qx},
    three effective diagrams for the
    self-energy corrections to the Wilson lines have been
    included in Eq.~(\ref{TFH1}).}
\begin{align}
H^{(\text{NLO})}_{\pi\gamma}(x,Q^2,k_T,\mu)
=
h_{\pi\gamma}(x,Q^2,k_T,\mu)
H^{(\text{LO})}_{\pi\gamma}(x,Q^2,k_T)
,\label{TFH1}
\end{align}
with the NLO correction function
\begin{align}
h_{\pi\gamma}(x,Q^2,k_T,\mu) = \frac{\alpha_s(\mu)C_F}{4\pi}\bigg\{
& -3\ln\frac{\mu^2}{Q^2}-\ln^2\frac{|k_T^2-xQ^2|}{Q^2}
+2\left[1-i\pi-i\pi\Theta\left(k_T^2-xQ^2\right)\right]\ln\frac{|k_T^2-xQ^2|}{Q^2}
\nonumber\\& -2\ln x
+\left(4\pi^2-i\pi\right)\Theta\left(k_T^2-xQ^2\right)-3-i5\pi
\bigg\}, \label{TFNLO}
\end{align}
$C_F$ being the color factor. The imaginary pieces proportional to
the step function $\Theta$ are generated from the $O(\alpha_s)$
quark diagrams. For the evaluation of the $O(\alpha_s)$ effective
diagrams, we have chosen the direction $n^\mu$ of the Wilson lines
the same as in \cite{Nandi:2007qx}
in order to respect the universality of the meson wave function.
Equation~(\ref{TFH1}) can also be achieved by substituting
$(Q^2+i\varepsilon)$ for the virtuality of the external photon, and
$(xQ^2-k_T^2+i\varepsilon)$ for the internal quark in
\cite{Nandi:2007qx}, and then employing the relations
$\ln(-Q^2-i\varepsilon) = \ln Q^2 -i\pi$ and
$\ln(-xQ^2+k_T^2-i\varepsilon) = \ln\left|xQ^2-k_T^2\right|- i\pi
\Theta(xQ^2-k_T^2)$.

Fourier transforming Eq.~(\ref{TFH1}) to the $b$ space, we arrive at
the NLO $k_T$ factorization formula for the time-like pion
transition factor
\begin{align}
    F_{\pi\gamma}^{\text{(NLO)}}(Q^2) =
i\pi & \frac{\sqrt{2}f_\pi}{6} \int_0^1 dx
\int_0^\infty b db\, \phi_\pi(x) 
\exp[-S(x,b,Q,\mu)]
\nonumber\\
&\times
\frac{\alpha_s(\mu)C_F}{4\pi}
\left[
~\widetilde{h}_{\pi\gamma}(x,Q^2,k_T,\mu)
H^{(1)}_0\left( \sqrt{x}Qb \right)
+ H^{(1)\prime \prime}_0\left( \sqrt{x}Qb \right)
\right],
\label{TFI1}
\end{align}
with
\begin{align}
\widetilde{h}_{\pi\gamma}(x,Q^2,k_T,\mu)
=
&
-3\ln\frac{\mu^2}{Q^2} - {1 \over 4}\ln^2 \frac{4x}{Q^2 b^2}
+ (1+\gamma_E-i{3\pi\over 2}) \ln \frac{4x}{Q^2 b^2} -2\ln x
\nonumber\\&
+{17 \pi^2 \over 12}+\pi-3-2 \gamma_E-\gamma_E^2 -i(4-3\gamma_E)\pi,
\label{TFh1}
\end{align}
$\gamma_E$ being the Euler constant. The function
\begin{align}
H^{(1)\prime \prime}_0\left( \rho \right)
\equiv \left[\frac{\partial^2}{\partial\alpha^2}H^{(1)}_\alpha(\rho )\right]_{\alpha=0} ,
\end{align}
where $\alpha$ denotes the order parameter of the Hankel function,
comes from the Fourier transformation of
$\ln^2(-xQ^2+k_T^2-i\varepsilon)$ in Eq.~(\ref{TFNLO}). For a small
argument $\rho = \sqrt{x}Qb \to 0$, its magnitude behaves as
$|H^{(1)\prime \prime}_0(\rho)| \sim
(1/3)\ln^2\rho~|H^{(1)}_0(\rho)|$, which represents a
double-logarithmic correction essentially. The perturbative
expansion could be improved by summing the double logarithm
$\alpha_s \ln^2 [x/(Q^2b^2)]$ in Eq.~(\ref{TFh1}), which arises from the
Fourier transformation of the term $\alpha_s\ln^2(|k_T^2-xQ^2|/Q^2)$ in
Eq.~(\ref{TFNLO}). Strictly speaking, it differs from the threshold
resummation of $\alpha_s\ln^2 x$ performed in \cite{Li:2001ay}, and
deserves a separate study. Besides, there is no end-point
enhancement involved in the present calculation, so we shall not
perform the resummation here for simplicity. Equations~(\ref{TFI0}) and
(\ref{TFI1}) will be investigated numerically in Sec.~\ref{Numeric}.

\begin{figure}[t]
    \begin{center}
         \includegraphics[height=5.5cm]{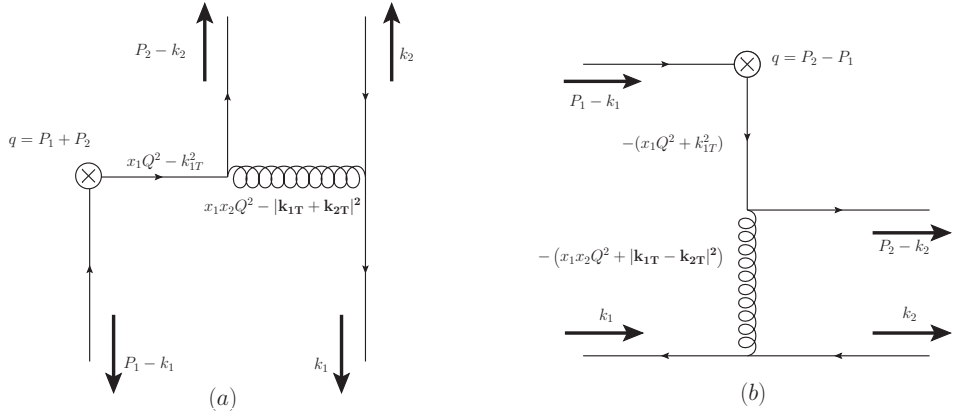}
    \end{center}
    \caption{LO quark diagrams for time-like and space-like pion electromagnetic form factors.}
    \label{fig:EMLO}
\end{figure}
\section{Pion Electromagnetic form factor}\label{PionEM}

We then derive the NLO, i.e., $O(\alpha_s^2)$ contribution to the
time-like pion EM form factor at leading twist. An LO quark diagram
for the corresponding scattering $\gamma^*(q) \to
\pi^+(P_1)~\pi^-(P_2)$ is depicted in Fig.~\ref{fig:EMLO}(a). We
choose light-cone coordinates, such that the momenta $P_1$ and $P_2$ are
parameterized the same as in Eq.~(\ref{TFcor}) with $Q^2 =q^2=
(P_1+P_2)^2>0$. The valence quark carries the momentum
$k_1=(x_1P_1^+,0,{\bf k}_{1T})$ and the valence anti-quark carries
$k_2=(0,x_2P_2^-,{\bf k}_{2T})$. The LO hard kernel reads
\begin{align}
    H^{(\text{LO})}_{\text{II}}(x_1,k_{1T},x_2,k_{2T},Q^2)
    = i4 \pi\alpha_s C_F\frac{x_1 \operatorname{Tr}
    \left[\feyn P_2\feyn P_1\gamma_\mu\feyn P_1\right]}{
    (x_1Q^2-{\bf k}_{1T}^2+i\varepsilon)
    (x_1 x_2Q^2-|{\bf k}_{1T}+{\bf k}_{2T}|^2+i\varepsilon)} ,
\label{EMH02}
\end{align}
where the denominators $(x_1Q^2-{\bf k}_{1T}^2)$ and $(x_1
x_2Q^2-|{\bf k}_{1T}+{\bf k}_{2T}|^2)$ are the virtuality of the
internal quark and gluon, respectively. The subscript $\text{II}$
denotes that the $k_T$-dependent terms in both the internal quark
and gluon propagators are retained. When one of the internal
particle propagators goes on mass shell, an imaginary part is
produced according to the principle-value prescription in
Eq.~(\ref{eq:Pri}).

Fourier transforming Eq.~(\ref{EMH02}) from the transverse-momentum
space $({\bf k}_{1T}, \,{\bf k}_{2T})$ to the impact-parameter space
$({\bf b}_1, \,{\bf b}_2)$, we obtain a double-$b$ convolution for
the LO time-like pion EM form factor \cite{Chen:2009sd}
\begin{align}
F_{\text{EM}}^{(\text{LO})}(Q^2)
    &= \frac{\pi^3 f_\pi^2 C_F}{2N_c} Q^2
\int_0^1dx_1 dx_2 \int_0^\infty db_1 db_2 b_1 b_2
\, \alpha_s(\mu) x_1
\phi_\pi(x_1) \phi_\pi(x_2) \exp[-S_{\text{II}}(x_1, b_1, x_2, b_2,
Q, \mu)] \nonumber\\
&\times H_0^{(1)}(\sqrt{x_1 x_2}Qb_2)
\left[H_0^{(1)}(\sqrt{x_1}Qb_1)J_0(\sqrt{x_1}Qb_2)\Theta(b_1-b_2)
    +H_0^{(1)}(\sqrt{x_1}Qb_2)J_0(\sqrt{x_1}Qb_1)\Theta(b_2-b_1)
\right] ,
\label{EMI2}
\end{align}
with the Bessel function of the first kind $J_0$, and the Sudakov
exponent $S_{\text{II}}(x_1, b_1, x_2, b_2, Q,\mu)=S(x_1, b_1,
Q,\mu)+S(x_2, b_2, Q,\mu)$. The above expression can also be
obtained via analytical continuation of the space-like form factor
in Fig.~\ref{fig:EMLO}(b) to the time-like region.

The NLO hard kernel for the space-like pion EM form factor has been
computed as the difference between the one-loop QCD quark diagrams
and effective diagrams in \cite{Li:2010nn}. To simplify the
calculation, the hierarchy $x_1Q^2, x_2Q^2\gg x_1x_2Q^2,k_{T}^2$ has
been postulated, since the $k_T$ factorization applies to processes
dominated by small-$x$ contributions. Ignoring the transverse
momenta of the internal quarks, the LO hard kernel in
Eq.~(\ref{EMH02}) reduces to
\begin{align}
H^{(\text{LO})}_{\text{I}}
(x_1,k_{1T},x_2,k_{2T},Q^2) =
i4\pi \alpha_s C_F\frac{\operatorname{Tr}\left[
\feyn P_2\feyn P_1\gamma_\mu\feyn P_1\right]}
{ Q^2(x_1 x_2Q^2-|{\bf k}_{1T}+{\bf k}_{2T}|^2+i\varepsilon)}   .
\label{EMH01}
\end{align}
The Fourier transformation of the above expression leads to a
single-$b$ convolution \cite{Gousset:1994yh}
\begin{align}
F_{\text{I}}^{(\text{LO})}(Q^2) = i\frac{\pi^2 f_\pi^2 C_F}{N_c}
\int_0^1dx_1 dx_2 \int_0^\infty db b
\,\alpha_s(\mu) \phi_\pi(x_1) \phi_\pi(x_2)
\exp[-S_{\text{I}} (x_1, x_2, b, Q, \mu)]
H_0^{(1)}(\sqrt{x_1 x_2}Qb),
\label{EMI1}
\end{align}
with the simplified Sudakov exponent $S_{\text{I}} (x_1, x_2, b, Q,
\mu) \equiv S_{\text{II}} (x_1, b, x_2, b, Q, \mu) $. Comparing the
outcomes from Eqs.~(\ref{EMI2}) and (\ref{EMI1}), we can justify the
proposed hierarchical relation, and tell which particle propagator,
the internal quark or the internal gluon, provides the major source
of the strong phase.

Substituting $(Q^2+i\varepsilon)$ for the virtuality of the external
photon, and $(x_1 x_2Q^2-|{\bf k}_{1T}+{\bf
k}_{2T}|^2+i\varepsilon)$ for the internal gluon in
\cite{Li:2010nn}, we have the NLO hard kernel for the time-like pion
EM form factor
\begin{align}
H^{(\text{NLO})}_{\text{EM}}(x_1,k_{1T},x_2,k_{2T},Q^2,\mu) =
h_{\text{EM}}(x_1, x_2, \delta_{12},Q,\mu)
H^{(\text{LO})}_{\text{I}}(x_1,k_{1T},x_2,k_{2T},Q^2) ,
\label{EMH1}
\end{align}
with the NLO correction function
\begin{align}
h_{\text{EM}}(x_1, x_2, \delta_{12},Q,\mu) =
{ \alpha_s(\mu) C_F \over 4 \pi } \bigg[
&
-{3 \over 4}\ln { \mu^2 \over Q^2}
- {17 \over 4}\ln^2 x_1  + {27 \over 8} \ln x_1 \ln x_2
- {13 \over 8} \ln x_1 + {31 \over 16} \ln x_2
\nonumber\\
&
-\ln^2 \delta_{12}
+\left({17 \over 4} \ln x_1 + {23 \over 8} +i2\pi \right) \ln \delta_{12}
+{  \pi^2 \over 12} + {1 \over 2} \ln 2 + {53 \over 4}-i{3\pi\over 4}
\bigg]
,
\label{h_1}
\end{align}
and the notation
\begin{align}
\ln{\delta_{12}} \equiv
\ln{\frac{\left| {|{\bf k}_{1T}+{\bf k}_{2T}|}^2 - x_1 x_2 Q^2 \right|}{Q^2}}
+i\pi\Theta\left({|{\bf k}_{1T}+{\bf k}_{2T}|}^2 - x_1 x_2 Q^2 \right) .
\label{DEL12}
\end{align}

Fourier transforming Eq.~(\ref{EMH1}), we derive the $k_T$
factorization formula for the NLO contribution at leading twist
\begin{align}
    F^{(\text{NLO})}_{\text{EM}}(Q^2) = {} &
i\frac{\pi f_\pi^2 C_F^2}{4N_c} \int_0^1dx_1 dx_2 \int_0^\infty db b
\, \alpha_s^2(\mu)  \phi_\pi(x_1) \phi_\pi(x_2)
\exp[-S_{\text{I}} (x_1, x_2, b, Q, \mu)]
\nonumber \\
&\times \left[
    \,\widetilde{h}_{\text{EM}}(x_1, x_2, b, Q, \mu) \, H_0^{(1)}(\sqrt{x_1 x_2}Qb)
+H^{(1)\prime \prime}_0\left(\sqrt{x_1 x_2}Qb\right)
\right] ,
\label{EM1I1}
\end{align}
with the function
\begin{align}
\widetilde{h}_{\text{EM}}(x_1, x_2, b, Q, \mu) = ~&
-{3 \over 4}\ln { \mu^2 \over Q^2}
-{1\over 4}\ln^2 \frac{4 x_1 x_2}{Q^2 b^2}
+\left(
{17\over 8}\ln x_1 +{23\over 16}+\gamma_E+i{\pi \over 2}
\right)
\ln \frac{4 x_1 x_2}{Q^2 b^2}
\nonumber\\
&
- {17 \over 4}\ln^2 x_1  + {27 \over 8} \ln x_1 \ln x_2
- \left( {13 \over 8} +{17\gamma_E \over 4} -i{17 \pi\over 8} \right)\ln x_1
+ {31 \over 16} \ln x_2
\nonumber\\
&
- {\pi^2 \over 2} + (1-2\gamma_E)\pi
+{1 \over 2}\ln 2+ {53 \over 4}
- {23 \over 8}\gamma_E - \gamma_E^2
+ i\left({171\over 16}+\gamma_E\right)\pi
.
\label{th_1}
\end{align}
The perturbative expansion could be improved by organizing the double
logarithm $\alpha_s\ln^2 x_1$ in Eq.~(\ref{h_1}) into the threshold
resummation factor $S_t( x_1, Q^2 )$ \cite{Li:2001ay}. This double
logarithm, the same as analyzed in \cite{Li:2001ay}, appears in the loop
correction to the virtual photon vertex under
the hierarchical relation $x_1Q^2\gg k_{T}^2$ \cite{Li:2010nn}.
Because there is no end-point enhancement involved
at leading twist, we shall not perform the threshold resummation here.
However, the end-point enhancement exists in the two-parton twist-3 contribution,
for which $S_t$ will play a crucial role, and be implemented in
Sec.~\ref{Numeric}. We shall investigate the NLO effect at leading
twist in the time-like pion EM form factor based on
Eqs.~(\ref{EMI2}) and (\ref{EM1I1}).

\section{Numerical Analysis}\label{Numeric}

The numerical analysis is performed in this section, for which we
adopt the standard two-loop QCD running coupling constant
$\alpha_s(\mu)$ with the QCD scale $\Lambda_{\rm QCD}=0.2$ GeV, the
pion decay constant $f_\pi = 0.131$ GeV, the nonasymptotic
two-parton twist-2 pion DA
\begin{align}
    \phi_\pi(x) = 6 x (1-x) \left[1+a_2 C_2^{3/2}(1-2x)\right]  ,
    \label{NAD}
\end{align}
with the Gegenbauer coefficient $a_2 = 0.2$ being fixed by lattice
QCD \cite{Braun:2006dg}, and the Gegenbauer polynomial
$C_2^{3/2}(u)=(3/2)(5u^2 -1)$.

\begin{figure}[t]
    \begin{center}
        \includegraphics[height=5.5cm]{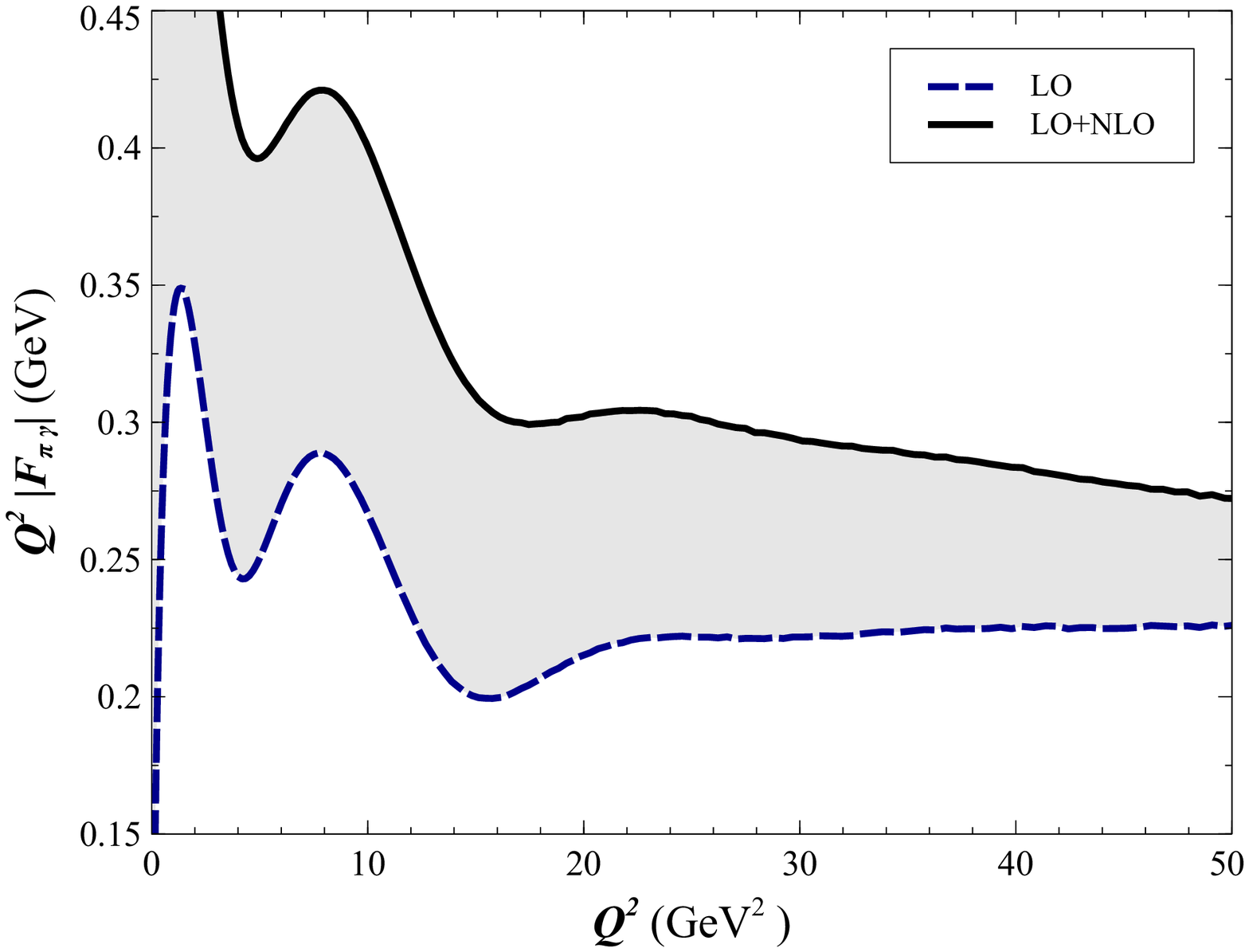}
        \includegraphics[height=5.5cm]{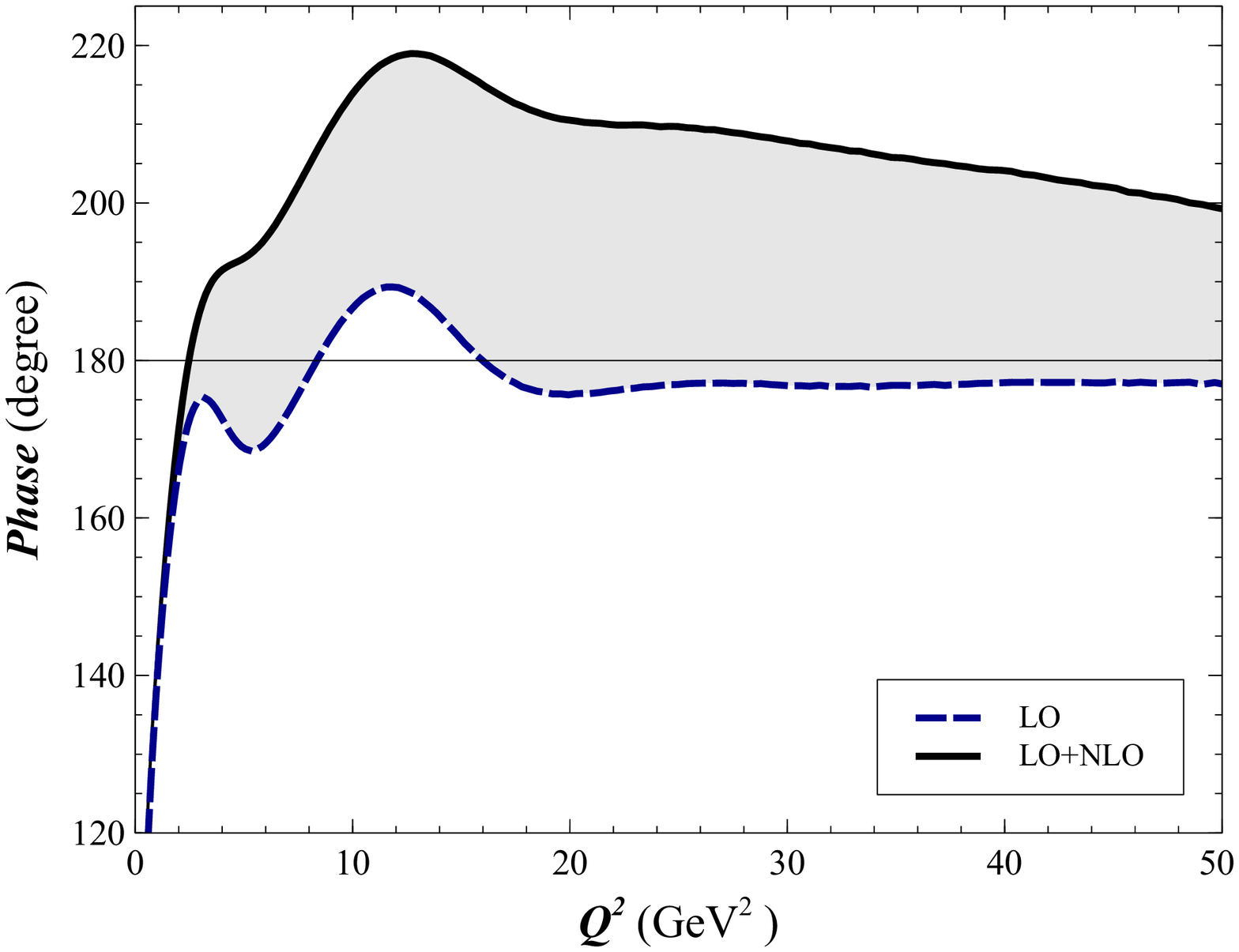}
    \end{center}
    \caption{
    Magnitude and phase of the
    time-like pion-photon transition form factor at LO (dashed) and
    up to NLO (solid). 
    The NLO correction is marked in gray.}
    \label{fig:TFNAD}
\end{figure}

We compute the LO and NLO contributions to the time-like pion-photon
transition form factor at leading twist via Eqs.~(\ref{TFI0}) and
(\ref{TFI1}), with the renormalization and factorization scale $\mu$
being set to the virtuality of the internal quark $\mu =
\max(\sqrt{x}Q, 1/b)$. The behavior of $Q^2 F_{\pi\gamma}(Q^2)$ for
$Q^2<20$ GeV\textsuperscript{2} displayed in Fig.~\ref{fig:TFNAD}
reflects the oscillatory nature of the LO hard kernel in the $b$
space. The LO time-like pion transition form factor exhibits an
asymptotic magnitude, $Q^2 \left|F_{\pi\gamma}(Q^2)\right| \approx
0.225$ GeV at large $Q^2$. Recall that an asymptotic scaling is
known as $Q^2 F_{\pi\gamma}(Q^2) \to \sqrt{2} f_\pi = 0.185$ GeV for
the space-like pion transition form factor at large $-q^2= Q^2$
\cite{BL}. The larger asymptotic value for the former is expected in
the $k_T$ factorization, because the internal quark may go on mass
shell for a time-like momentum transfer $q$, but it is always
off-shell for a space-like $q$. The ratio between the asymptotic
values of the above two transition form factors is roughly $1.22$,
comparable to the data $1.14$ from the $\eta - \gamma$ transition
form factors for $Q^2 >100$ GeV\textsuperscript{2}
\cite{Aubert:2006cy}. Note that the time-like and space-like
transition form factors would have equal magnitudes in the LO
collinear factorization without including the parton transverse
momentum $k_T$. The NLO contribution to the time-like pion
transition form factor is also displayed in Fig.~\ref{fig:TFNAD},
which decreases with $Q^2$ as expected in PQCD. Compared to the LO
result, the NLO correction to the magnitude is about $30\%$ at $Q^2
= 30$ GeV\textsuperscript{2}, and less than $20\%$ for $Q^2 > 50$
GeV\textsuperscript{2}.

For the phase, the LO result arises with $Q^2$, and approaches an
asymptotic value close to $180^\circ$ as shown in
Fig.~\ref{fig:TFNAD}. It is obvious that the variation with $Q^2$ is
also attributed to the inclusion of the parton transverse momentum
$k_T$. If $k_T$ in Eq.(\ref{TFH0}) was dropped, the LO hard kernel
reduces to the traditional expression in the collinear
factorization, which always leads to a real $F_{\pi\gamma}$. A
quantitative understanding can be attained via Eq.~(\ref{eq:Pri}):
the contributions from the two terms in Eq.~(\ref{eq:Pri}) are
comparable at low $Q^2$, such that the time-like pion transition
form factor acquires a nontrivial phase. At high $Q^2> 20$
GeV\textsuperscript{2}, the phase is dominated by the first term in
Eq.~(\ref{eq:Pri}), since it is unlikely to have a large parton
$k_T^2=xQ^2$ demanded by the second term. That is, the tiny
deviation (less than $5^\circ$) of the asymptotic phase from
$180^\circ$ is caused by the power-suppressed $k_T^2/Q^2$ effect.
The NLO correction to the phase is about $30^\circ$ at $Q^2 = 30$
GeV\textsuperscript{2}, and fewer than $20^\circ$ for $Q^2 > 50$
GeV\textsuperscript{2}. The above investigation implies that
higher-order corrections to the complex time-like transition form
factors are under control in the $k_T$ factorization. 
As stated before, the perturbative expansion could be improved by
resumming the double logarithm $\alpha_s \ln^2 x$ in
Eq.~(\ref{TFh1}).
%

\begin{figure}[t]
    \begin{center}
        \includegraphics[height=5.5cm]{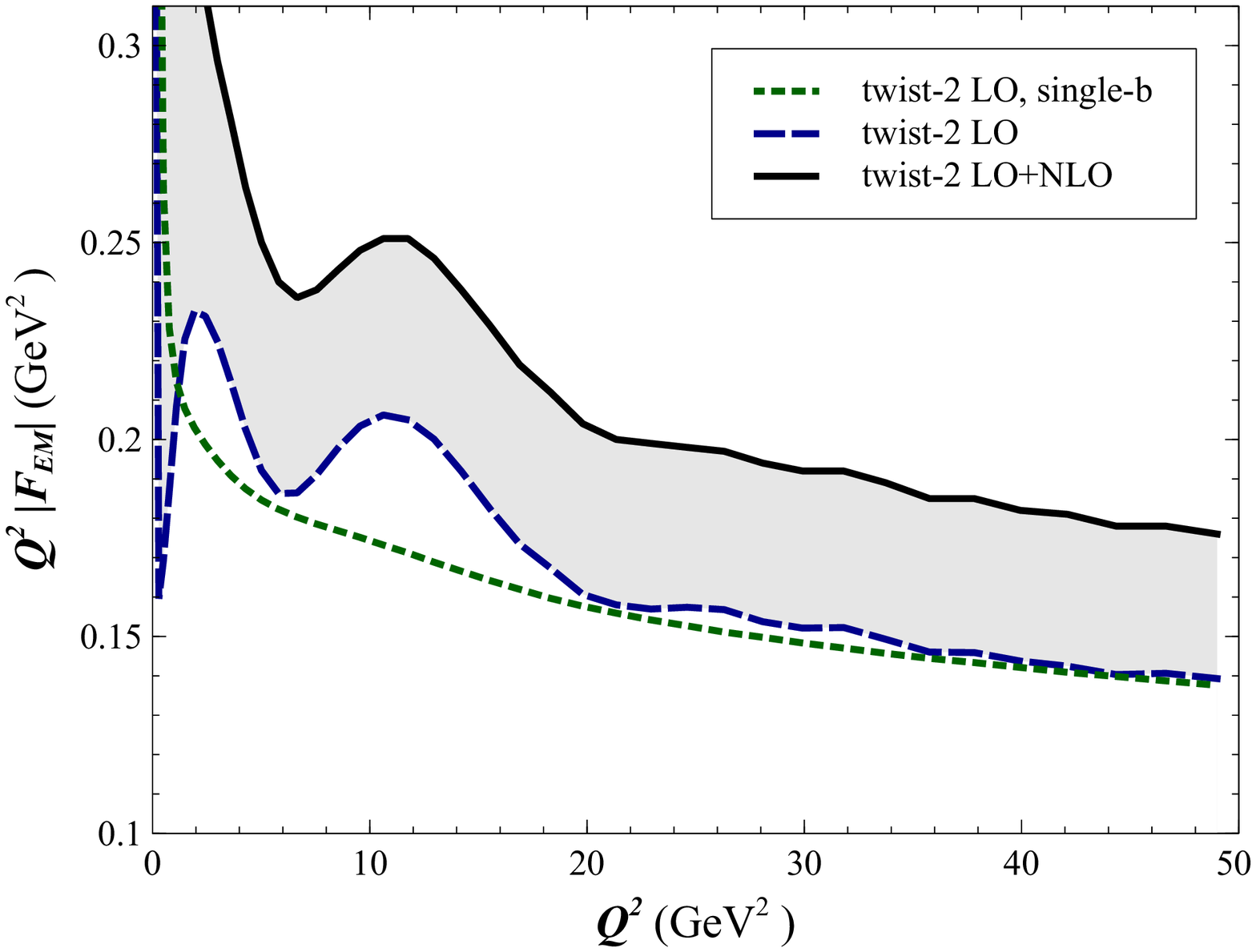}
        \includegraphics[height=5.5cm]{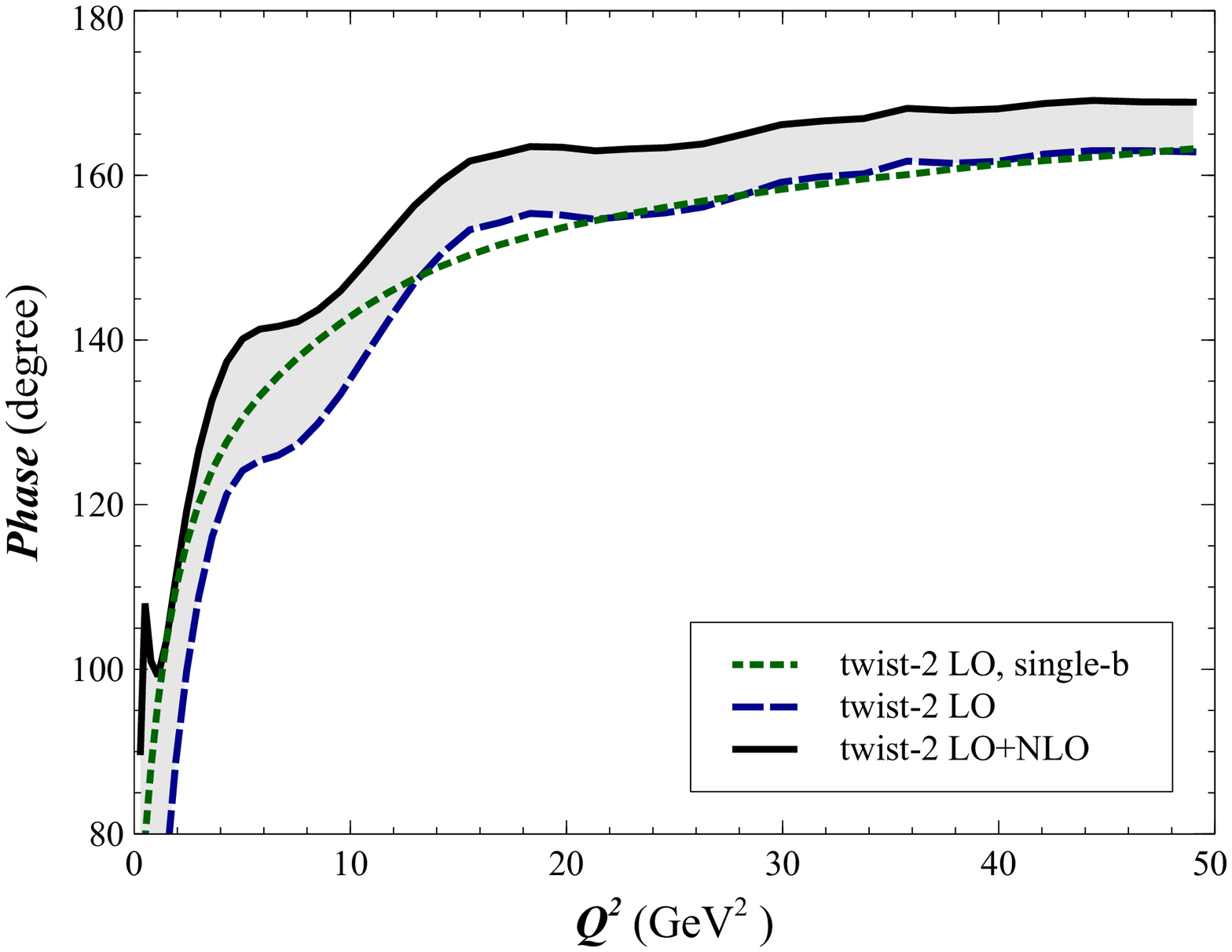}
    \end{center}
    \caption{Magnitude and phase of the time-like pion EM form factor at leading twist.
        Contributions from LO with the single-$b$ convolution (dotted),
        LO (dashed), and LO+NLO (solid) are shown.
    }
    \label{fig:EMtw2}
\end{figure}

For the analysis of the time-like pion EM form factor, we first
identify the major source of the strong phase by comparing the
results from Eqs.~(\ref{EMI2}) and (\ref{EMI1}) in
Fig.~\ref{fig:EMtw2}. The renormalization and factorization scale
$\mu$ is set to $\mu = \max\left( \sqrt{x_1}Q, 1/b_1, 1/b_2 \right)$
\cite{LS,Li:2010nn}, associated with the virtuality of the internal
particles. The curve from Eq.~(\ref{EMI2}) implies that the
magnitude of the time-like pion EM form factor has an asymptotic
behavior $Q^2 |F_{\text{EM}}(Q^2)| \to 0.14$ GeV\textsuperscript{2}
as $Q \to \infty$. Similar to the case of the transition form
factor, this asymptotic value is larger than that of the space-like
pion EM form factor \cite{Li:2010nn}, because of the inclusion of
the parton transverse momenta $k_T$. The inclusion of $k_T$ also
leads to the variation of the phase with $Q^2$, which arises from
the first quadrant, and then approaches an asymptotic value close to
$165^\circ$. For $Q^2>15$ GeV\textsuperscript{2}, the difference
between using the single-$b$ and double-$b$ convolutions is
insignificant in both magnitude and phase, verifying the
hierarchical relation $x_1Q^2, x_2Q^2\gg x_1x_2Q^2,k_{T}^2$, and the
major source for the strong phase as the internal gluon propagator.
We then investigate the NLO effect in the time-like pion EM form
factor based on Eqs.~(\ref{EMI2}) and (\ref{EM1I1}), which is also
shown in Fig.~\ref{fig:EMtw2}. For $Q^2 > 10$
GeV\textsuperscript{2}, the observed NLO correction is roughly
$25\%$ for the magnitude, and less then $10^\circ$ for the phase.
That is, the perturbative evaluation of the time-like pion EM form
factor is stable against radiative corrections at leading twist.

\begin{figure}[t]
    \begin{center}
        \includegraphics[height=5.5cm]{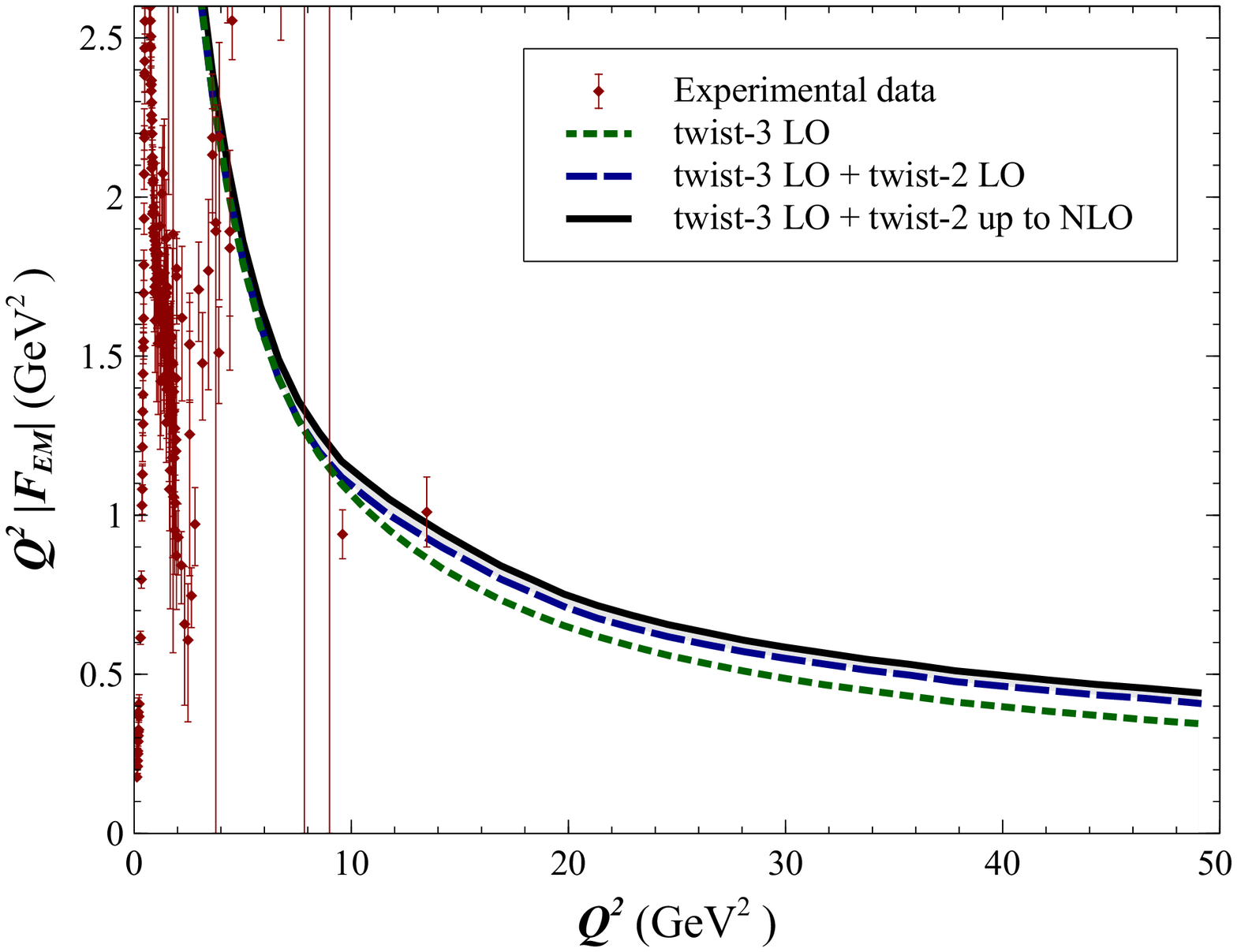}
        \includegraphics[height=5.5cm]{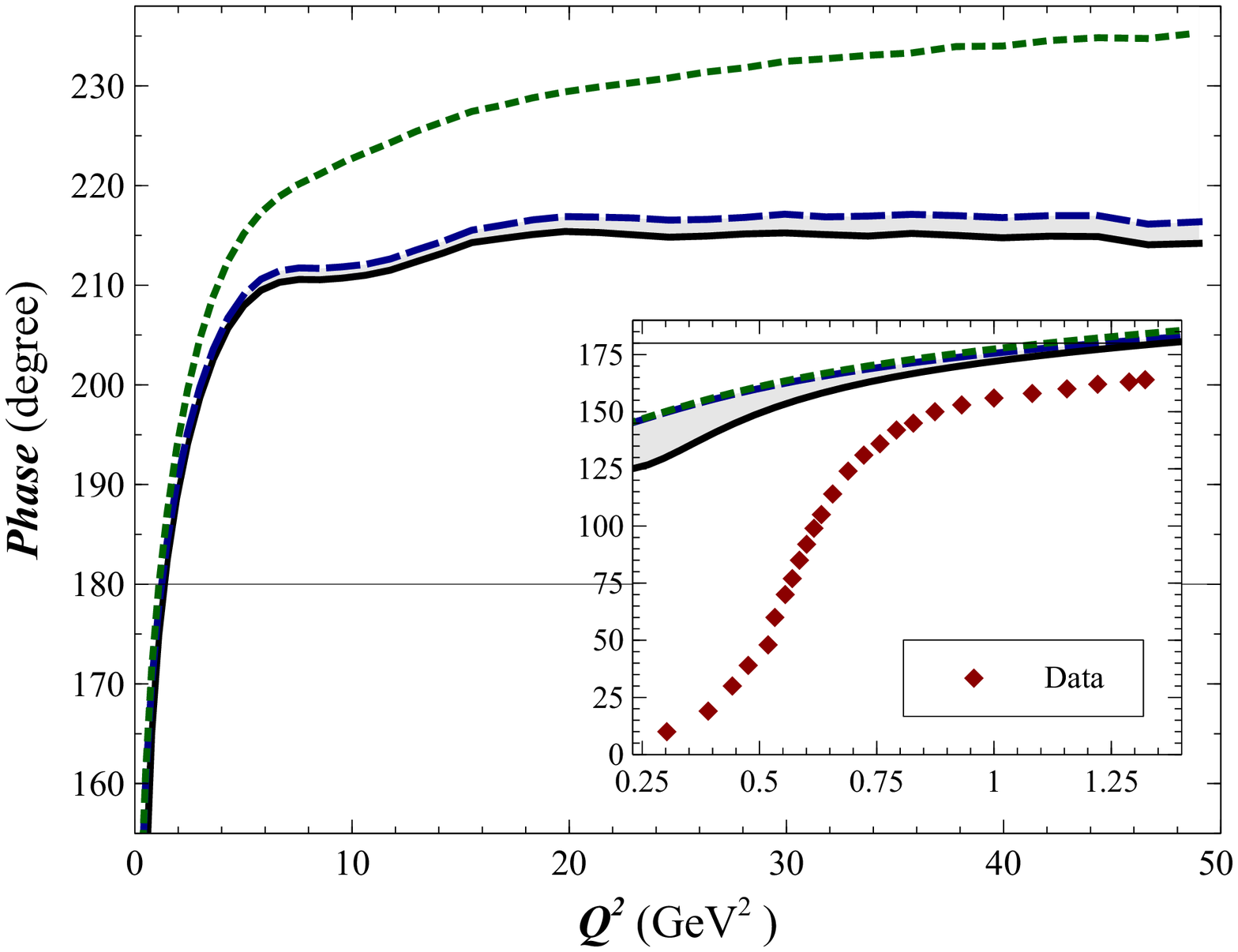}
    \end{center}
    \caption{
        Contributions to the time-like pion EM form factor from two-parton twist-3 LO (dotted),
        two-parton twist-3 LO plus twist-2 LO (dashed),
        and two-parton twist-3 LO plus twist-2 up to NLO (solid).}
    \label{fig:EMtw3}
\end{figure}

At last, we include another piece of subleading effects, the LO
two-parton twist-3 contribution \cite{Nagashima:2002iw}, for completeness.
Following the same derivation of the twist-2 contribution in Sec.~\ref{PionEM},
the $k_T$ factorization formula for the LO two-parton twist-3 contribution
to the time-like pion EM form factor was obtained in \cite{Chen:2009sd}, where
an explicit double-$b$ convolution expression similar to Eq.~(\ref{EMI2})
can be found. The hard kernel in the impact-parameter space is identical
to the one in Eq.~(\ref{EMI2}).
We employ the asymptotic two-parton twist-3 DAs,
\begin{align}
    \phi_\pi^P(x) = 1 ,
\;\;\;\;\;\;
    \phi_\pi^T(x) = 1-2x ,
    \label{Phi3}
\end{align}
with the associated chiral scale $\mu_\pi=1.3$ GeV. The threshold
resummation factor $S_t(x, Q)$ with a shape parameter $c = 0.4$ is
included, since the important double logarithm $\alpha_s \ln^2 x$ at
small $x$ needs to be summed \cite{Li:2001ay}. The numerical
outcomes for the time-like pion EM form factor are presented in
Fig.~\ref{fig:EMtw3}, where the available experimental data
\cite{Whalley2003aaa,Proto1973} are displayed for comparison. It is
known that the pion EM form factor is dominated by the two-parton
twist-3 contribution, instead of by the twist-2 one at currently
accessible energies, because of the end-point enhancement developed
by the above DAs \cite{Cao:1997st,Huang:2004su,Chen:2009sd}.
This enhancement is understood easily as follows:
the virtual quark and gluon propagators behave like $1/x_1$ and $1/(x_1 x_2)$,
respectively, as indicated in
Eq.~(\ref{EMH02}). The twist-2 pion DA is proportional to $\phi_\pi(x) \sim O(x)$,
but the twist-3 pion DAs remain constant $\phi^{P,T}_\pi(x) \sim O(1)$
at small $x$, which then enhance the end-point contribution
dramatically. This enhancement was also observed in perturbative evaluation
of the $B\to\pi$ transition form factors \cite{KLS01}, and confirmed by the
light-cone sum-rule analysis \cite{KP98}.
The relative phase between the twist-2 and two-parton twist-3 pieces is
about $70^\circ$ as indicated by Figs.~\ref{fig:EMtw2} and
\ref{fig:EMtw3}, so the magnitude of the form factor is hardly
affected by the former. However, the twist-2 contribution does have
a sizable effect on the phase as illustrated in
Fig.~\ref{fig:EMtw3}.

The predictions for the magnitude of the time-like pion EM form
factor from the $k_T$ factorization can accommodate the data
\cite{Whalley2003aaa} for $Q^2>4$ GeV\textsuperscript{2}, an
observation consistent with that from the LO analysis
\cite{Chen:2009sd}. We point out that the measured magnitude of the
time-like pion EM form factor is larger than the space-like one
\cite{Chen:2009sd}, and simultaneous accommodation of both data is
possible in the $k_T$ factorization, but not in the collinear
factorization. Though the perturbative calculations may not be
justified for small $Q^2 < 4$ GeV\textsuperscript{2}, it is
interesting to see the coincidence between the increases of the
phase with $Q^2$ from the $k_T$ factorization and from the data for
$Q^2<1.3$ GeV\textsuperscript{2}. In a Breit-Wigner picture, the
observed phase increase could be attributed to a resonant $\rho$
meson propagator \cite{Proto1973}. It happens that the parton
transverse momentum $k_T$ plays the role of the $\rho$ meson mass,
such that the two curves in Fig.~\ref{fig:EMtw3} exhibit the similar
tendency, and begin to merge for $Q^2 > 1$ GeV\textsuperscript{2}.
Again, this coincidence cannot be achieved in the collinear
factorization, which does not generate a significant phase shift.

The consistency between the present analysis and the data supports
the $k_T$ factorization formalism
as an appropriate framework for
studying complex time-like form factors. It has been understood that
the complex penguin annihilation contribution is essential for
explaining direct CP asymmetries in two-body hadronic $B$ meson
decays \cite{KLS}. This contribution involves time-like scalar form
factors, which can be calculated in the same $k_T$ factorization
formalism. It has been observed that the phase of the $S$-wave
component in $\pi\pi$ scattering shows a similar $Q^2$ dependence to
that of the $P$-wave \cite{Proto1973}. Therefore, the PQCD
predictions for the above direct CP asymmetries are expected to be
reliable. The formalism for three-body hadronic $B$ meson decays
\cite{CL03} has required the introduction of two-meson wave
functions, whose parametrization also involves time-like form
factors of various currents. Stimulated by our work, we have the
confidence on computing these complex time-like form factors
directly in the PQCD approach.


\section{Conclusions}\label{Concl}

In this Letter we have calculated the time-like pion-photon
transition and EM form factors up to NLO in the $k_T$ factorization
formalism. The corresponding NLO hard kernels were derived by
analytically continuing the space-like ones to the time-like region
of the momentum transfer squared $Q^2$. We have identified the
$k_T$-dependent internal gluon propagator as the major source for
the strong phase of the time-like pion EM form factor, which
increases with $Q^2$, and approaches an asymptotic value
\cite{Ananthanarayan:2012tn}. The magnitudes of the time-like form
factors are larger than those of the space-like ones. It has been
realized that the above features are attributed to the inclusion of
the parton transverse momenta, and consistent with the tendency
implied by the data. It was observed that the NLO corrections in
magnitude (phase) change the LO leading-twist results by roughly
$30\%$ ($30^\circ$) for the pion transition form factor, and $25\%$
($10^\circ$) for the pion EM form factor as $Q^2> 30$ GeV$^2$. The
stability against radiative corrections justifies the $k_T$
factorization formalism for both time-like form factors at leading
twist. Therefore, the predictions for strong phases of annihilation
contributions to two-body hadronic $B$ meson decays in the PQCD
approach may be reliable. The framework presented here will have
other applications, for example, to the construction of the
two-meson wave functions for three-body $B$ meson decays.


\vskip 0.3cm We thank B. Ananthanarayan and I. Caprini for useful discussions.
The work was supported in part by the National Science
Council of R.O.C. under Grant No. NSC-98-2112-M-001-015-MY3, and by
the National Center for Theoretical Sciences of R.O.C.



\begin{thebibliography}{99}
%
\bibitem{CCH} S.~Catani, M.~Ciafaloni and F.~Hautmann, Phys. Lett. B {\bf 242},
97 (1990); Nucl. Phys. B {\bf 366}, 135 (1991).
\bibitem{CE} J.C.~Collins and R.K.~Ellis, Nucl. Phys. B {\bf 360}, 3 (1991).
\bibitem{LRS} E.M.~Levin, M.G.~Ryskin, Yu.M.~Shabelskii, and A.G.~Shuvaev,
Sov. J. Nucl. Phys. {\bf 53}, 657 (1991).
\bibitem{BS} J.~Botts and G.~Sterman, Nucl. Phys. B {\bf 325}, 62 (1989).
\bibitem{LS} H.-n.~Li and G.~Sterman, Nucl. Phys. B {\bf 381}, 129 (1992).
\bibitem{HS} T.~Huang and Q.X.~Shen, Z. Phys. C {\bf 50}, 139 (1991);
J.P.~Ralston and B.~Pire, Phys. Rev. Lett. {\bf 65}, 2343 (1990).
%
\bibitem{Nandi:2007qx}
S.~Nandi and H.-n.~Li,
Phys. Rev. D {\bf 76}, 034008 (2007).
%
\bibitem{Li:2009pr}
H.-n.~Li and S.~Mishima,
Phys. Rev. D {\bf 80}, 074024 (2009).
%
\bibitem{Li:2010nn}
H.-n.~Li, Y.-L.~Shen, Y.-M.~Wang, and H.~Zou.
Phys. Rev. D {\bf 83}, 054029 (2011).
%
\bibitem{Li:2012nk}
H.-n.~Li, Y.-L.~Shen, and Y.-M.~Wang,
arXiv:1201.5066 [hep-ph] (2012).
%
\bibitem{KLS} Y.Y.~Keum, H-n.~Li, and A.I.~Sanda,
Phys. Lett. B {\bf 504}, 6 (2001); Phys. Rev. D {\bf 63}, 054008 (2001);
Y.Y.~Keum and H.-n.~Li, Phys. Rev. D {\bf 63}, 074006 (2001);
C.D.~Lu, K.~Ukai and M.Z.~Yang, Phys. Rev. D {\bf 63}, 074009 (2001).
%
\bibitem{CL03} C.H.~Chen and H.-n.~Li, Phys. Lett. B {\bf 561}, 258 (2003);
Phys. Rev. D {\bf 70}, 054006 (2004).
\bibitem{MP} D.~Muller et~al., Fortschr. Physik. {\bf 42}, 101 (1994);
M.~Diehl, T.~Gousset, B.~Pire, and O.~Teryaev, Phys. Rev. Lett. {\bf
81}, 1782 (1998); M.V.~Polyakov, Nucl. Phys. {\bf B555}, 231 (1999).
%
\bibitem{Jakob:1994hd}
  R.~Jakob, P.~Kroll and M.~Raulfs,
  J.\ Phys.\ G {\bf 22}, 45 (1996).  
%
\bibitem{Nagashima:2002ia}
M.~Nagashima and H.-n.~Li,
Phys. Rev. D {\bf 67}, 034001 (2003).
%
\bibitem{Li:2000hh}
H.-n.~Li.
Phys. Rev. D {\bf 64}, 014019 (2001)
%
\bibitem{Jakob}
    R.~Jakob and P.~Kroll, Phys. Lett. B {\bf 315}, 463 (1993);
    B {\bf 319}, 545 (1993).
%
\bibitem{Brodsky80}
S.~J. Brodsky, T. Huang, and G.~P. Lepage, SLAC-PUB-2540;
S.J. Brodsky, P. Hoyer, C. Peterson, and N. Sakai, Phys. Lett. B
{\bf 93}, 451 (1980);
S.J. Brodsky, C. Peterson, and N. Sakai, Phys.
Rev. D {\bf 23}, 2745 (1981).
%
\bibitem{Raha:2010kz}
U.~Raha and H.~Kohyama, Phys. Rev. D {\bf 82}, 114012 (2010).
%
\bibitem{Li:2001ay}
H.-n.~Li,
Phys. Rev. D {\bf 66}, 094010 (2002);
K.~Ukai and H.-n.~Li,
Phys. Lett. B {\bf 555}, 197 (2003).
%
\bibitem{Li:1994zm}
H.-n.~Li.
Phys. Rev. D {\bf 52}, 3958 (1995).
%
\bibitem{Chen:2009sd}
J.-W.~Chen, H.~Kohyama, K.~Ohnishi, U.~Raha, and Y.-L.~Shen,
Phys. Lett. B {\bf 693}, 102 (2010).
%
\bibitem{Gousset:1994yh}
T.~Gousset and B.~Pire,
Phys. Rev. D {\bf 51}, 15 (1995).
%
\bibitem{Braun:2006dg}
V.~Braun, M.~Gockeler, R.~Horsley, H.~Perlt, D.~Pleiter, et~al.,
Phys. Rev. D {\bf 74}, 074501 (2006).
%
\bibitem{BL}
S.J.~Brodsky and G.P.~Lepage, Phys. Rev. D {\bf 24}, 1808 (1981).
%
\bibitem{Aubert:2006cy}
Bernard~Aubert et~al.
Phys. Rev. D {\bf 74}, 012002 (2006).
%
%
\bibitem{Nagashima:2002iw}
M.~Nagashima and H.-n.~Li,
Eur. Phys. J. C {\bf 40}, 395 (2005).
\bibitem{Whalley2003aaa}
M.R.~Whalley,
J. Phys. G {\bf 29}, A1 (2003);
%
J.~Milana, S.~Nussinov, and M.G.~Olsson,
Phys. Rev. Lett. {\bf 71}, 2533 (1993);
%
T.K.~Pedlar et~al.,
Phys. Rev. Lett. {\bf 95}, 261803 (2005).
\bibitem{Proto1973}
S.D.~Protopopescu, M.~Alston-Garnjost, A.~Barbaro-Galtieri et al.,
Phys. Rev. D {\bf 7}, 1279 (1973).
%
\bibitem{Cao:1997st}
F.-g.~Cao, Y.-b.~Dai, and C.-s.~Huang,
Eur. Phys. J. C {\bf 11}, 501 (1999).
%
\bibitem{Huang:2004su}
T.~Huang and X.-G.~Wu,
Phys. Rev. D {\bf 70}, 093013 (2004).
%
\bibitem{KLS01}
T.~Kurimoto, H.-n.~Li, and A.~I.~Sanda, Phys. Rev. D {\bf 65}, 014007
(2002).
%
\bibitem{KP98}
A.~Khodjamirian, R.~Ruckl, and C.~W.~Winhart, Phys. Rev. D 58, 054013 (1998).
%
\bibitem{Ananthanarayan:2012tn}
  B.~Ananthanarayan, I.~Caprini and I.~S.~Imsong,
  Phys.\ Rev.\ D {\bf 85}, 096006 (2012).  

\end{thebibliography}
\end{document}